\documentstyle{cupconf}
\title[Far-UV Radiation from Galaxies]{The Far-Ultraviolet Radiation from \\
Elliptical Galaxies}

\author[Ben Dorman]%
{B\ls E\ls N\ns D\ls O\ls R\ls M\ls A\ls N}

\affiliation{Laboratory for Astronomy \& Solar Physics, \\
	NASA/GSFC, Greenbelt, MD 20771 \\ and\\
	Department of Astronomy, University of Virginia,\\
Charlottesville, VA 22903-0818, USA}

\begin{document}

\def\mathmode#1{\ifmmode {#1} 
                  \else {$#1\mkern-5mu$} \fi}

\def\mlam{\mathmode{m_\lambda}}
\def\xvv{\hbox{$15\!-\!V$}}
\def\ngc#1{\hbox{\rm NGC\thinspace#1}}
\def\et{et~al.}
\def\teff{\mathmode{T_{_{\rm eff}}}}
\def\wrange#1#2{\mathmode{ {#1} < \lambda < {#2} \, {\rm \AA}}}
\def\wless#1{\hbox{\mathmode{\lambda < {#1} \, {\rm \AA}}}}
\def\wmore#1{\hbox{\mathmode{\lambda > {#1} \, {\rm \AA}}}}
\def\gta{\mathmode{\gtrsim}}
\def\lta{\mathmode{\lesssim}}
\def\mgii{\mathmode{\rm Mg_2}}
\def\xvv{\mathmode{(15-V)}}
\def\solar{\mathmode{_\odot}}
\def\etal{{\it et~al.}}
\def\kelvin{\mathmode{\, {\rm K}}}
\def\lta{\hbox{$\;$\raise 0.6 ex \hbox{$<$}\kern -1.8 ex\lower .5 
ex\hbox{$\sim$}$\;$}}
\def\gta{\hbox{$\;$\raise 0.6 ex \hbox{$>$}\kern -1.7 ex\lower .5 
ex\hbox{$\sim$}$\;$}}
\def\mc{\mathmode{M_c}}
\def\lv{\mathmode{L_V}}
\def\lbol{\mathmode{L_{\rm bol}}}
\def\lvs{\mathmode{L_V^\odot}}
\def\Lv{\mathmode{{\cal L}_V}}
\def\mass{\mathmode{m^\prime}}
\def\Lvo{\mathmode{{\cal L}^{\rm E}_{\rm V} }}
\def\Lband{\mathmode{{{\cal L}_\lambda}}}
\def\Lbando{\mathmode{{\cal L}^{\rm E}_\lambda}}
\def\Lbandl{\mathmode{{\cal L}^{\rm L}_\lambda}}
\def\Ebandl{\mathmode{{\cal E}^{\rm L}_\lambda}}
\def\eband{\mathmode{E_\lambda}}
\def\lband{\mathmode{L_\lambda}}
\def\sst{\scriptscriptstyle}

\def\lbando{\mathmode{L_\lambda^{\sst E}}}
\def\lbandl{\mathmode{L_\lambda^{\sst L}}}
\def\sbprate{\mathmode{{\it {\dot n}}_0(t)}}
\def\fuv{\mathmode{{\it f}_{\sst H}}}
\def\fred{\mathmode{{\it f}_{\sst C}}}
\def\Ered{\mathmode{{{\cal E}_{_{V}}^{^{\rm red}}}}}
\def\Euv{\mathmode{{{\cal E}_\lambda^{^{\rm UV}}}}}
\def\bclam{\mathmode{\alpha_\lambda}}
\def\exv{ \mathmode{E_{1500}} }
\def\exxv{\mathmode{E_{2500}}}
\def\exxxiii{\mathmode{E_{3300}}}
\def\ev{\mathmode{E_V}}
\def\EVcool{\mathmode{{{\cal E}_{_{V}}^{{^{C}}}}}}
\def\ELcool{\mathmode{{{\cal E}_{\lambda}^{^{C}}}}}
\def\ELhot{\mathmode{{{\cal E}_{\lambda}^{^{H}}}}}
\def\EVhot{\mathmode{{{\cal E}_{_{V}}^{^{H}}}}}
\def\Exvhot{\mathmode{{{\cal E}_{1500}^{^{H}}}}}
\def\Exvcool{\mathmode{{{\cal E}_{1500}^{^{C}}}}}

\def\ibhb{IBHB}
\def\menv{\mathmode{M_{\rm env}}}
\def\FUV{\mathmode{F_{\sst UV}}}

\makepptitle

\begin{abstract}

Since the discovery of the Ultraviolet Upturn Phenomenon
(``UVX'') in early-type galaxies it has been clear that the stellar
populations
of such systems contain an unexpected hot component. Recent work has
provided strong circumstantial evidence that the stars radiating 
at short wavelengths (\wless{2000}) is in fact due to hot horizontal
branch, post-HB stars and post-AGB stars. We summarize the arguments in
favour of this hypothesis. We then derive an estimate for the fraction of
all HB stars that must be contributing to the UV upturn phenomenon in the
strongest UVX galaxy, NGC 1399, and derive a hot star
fraction $f_{H} \sim 0.16.$ The implication
is that UVX arises from a minority fraction of the dominant stellar
population. We conclude that the mechanism that produces the UVX is not
one that can be explained naturally by the presence of an extremely metal-rich
or metal-poor population.

\end{abstract}

\firstsection 
\section{Introduction}

The Ultraviolet Upturn Phenomenon was first found by the OAO-2
spacecraft in the late sixties (\cite{co69}).
It consists of a `UV rising
branch' at wavelengths shorter than \wless{2500}, which varies in
amplitude amongst the galaxies observed. The amplitude of the UVX phenomenon
varies by a factor of about 10 from NGC 1399 [$\xvv = 2.05$] to
M32 [$\xvv = 4.50$] with very similar slopes in the IUE spectral
range (\wless{1200}).The question was investigated
in further detail by Burstein \etal\  (1988, hereafter B3FL)
who presented a sample of
galaxies observed by the International Ultraviolet Explorer (IUE).
They  determined  the UV/Optical color $\mlam(1550) - V$ [hereafter 
\xvv] from ground-based photometry. They
plotted these colors against the absorption line index \mgii, a
reliable metallicity indicator for
the Galactic globular cluster system (\cite{bfgk}; \cite{bh90}),  and
the central velocity dispersion $\sigma_0.$  For the galaxies that 
are quiescent, i.e. which contain no evidence of recent star formation and
 nuclear activity  there is a non-linear  correlation between
 \mgii\  and the UVX amplitude (see also \cite{f83}).
 The UVX was also found to
correlate, albeit not as tightly, with $\sigma_0,$ 
which is an indicator of luminosity through the Faber-Jackson
relation.
Thus either the amplitude increase is directly related to
the metallicity (or at least magnesium  abundance) of the galaxies, or there
 is a tendency for bright galaxies to have strong UV upturns.
Enhancements in $\alpha$-capture elements such as Mg  have been
found to be a feature of the spectra of bright ellipticals
(\cite{wfg92}) and imply something about the early enrichment
history of massive systems.

This contribution is organized as follows: in the next section we discuss
the recent observational evidence pertaining to the UVX.
Section 3 summarizes the problems associated with modelling the
Horizontal Branch. In \S 4 we give an overview of evolutionary
population synthesis, and use it to derive estimates of the
fraction of hot stars present as a function of the UV upturn strength.
\S 5 discusses these estimates in the light of population models
for the UVX.

\section{The `Old Population' Model} 

\subsection{UV sources in Old Populations}

The potential sources of UV radiation in old stellar populations have
been reviewed in detail by \cite{gr90}, to which the reader is referred
for a full discussion. Briefly, the sources that are likely to produce
the largest contribution to the UV output from a galaxy are evolved
stars powered by stable nuclear reactions.  The production of a hot
star on the HB requires that much or all of the hydrogen rich envelope
has been stripped from the star prior to arrival on the HB. The sources
are (see Dorman, Rood, \& O'Connell 1993, hereafter \cite{dro93} and
references therein):

\begin{itemize}

\item Post-Asymptotic Giant
Branch (P-AGB) stars, which are bright ($L/\lsun \sim 1,000 - 10,000$)
but short-lived $(t \la 25,000\,{\rm yr})$ . These  are
likely the most common `exit channel' for metal-rich stars dying as 
white dwarfs after going through thermal pulses on the AGB. 

\item Post-Early Asymptotic Giant Branch stars, which miss the thermally-
pulsing stage of AGB evolution, and cross the HR diagram at $L/\lsun \sim 
1,000$ in $\sim 10^6 \, {\rm yr}.$

\item Hot Post-HB stars, termed AGB-Manqu\'e stars, which 
are intermediate in brightness ($L/\lsun \sim 100 - 500$) and
lifetime ($t \sim 2-4 \times 10^7 \, {\rm yr}).$

\item Hot Horizontal Branch (HB) stars, which are fainter
($L/\lsun \sim 10 - 50$) but long-lived ($t  \sim 10^8 \, {\rm yr}$).
For convenience, these are grouped as either ``Intermediate Blue 
HB stars'' (IBHB) if their envelopes are large enough to allow ascent up
the AGB, and ``Extreme HB stars'' (EHB) if not.

\end{itemize} 

All of these sources have their observational counterparts in the
Galactic field or in globular clusters (see \cite{dor95} and Dorman
1997a,b for summaries). P-AGB stars are present as UV-bright stars or
the nuclei of planetary nebul\ae. Their total UV lifetime radiation
(and contribution to the integrated spectrum) in this phase decreases
strongly with increasing luminosity, since both the total fuel
available to them decreases and its rate of consumption increases.  It is
well-known that hot HB stars are produced in large numbers in
metal-poor globular clusters. The Galactic field hot subdwarfs
(designated sdB, sdO and variants; see \cite{gs74}; \cite{sa94})
dominate the UV excess sources in surveys of our Galaxy (\cite{pg} and
others). They are kinematically related to the old disk (\cite{sl95})
which is predominantly metal-rich compared to the halo clusters. 
The sdB stars occupy the 
same HR diagram location as the EHB stars. The AGB-Manqu\'e
stars, progeny of the EHB population, are similarly associated with the
sdO population.  

The elliptical galaxies are dominated by metal-rich populations (probably of 
solar metallicity or above). Until recently, the
only evidence for hot HB stars in a metal-rich environment was in the
Galactic field. There were no unambiguously metal-rich populations
that contained them i.e. in the coeval, homogeneous Galactic clusters.  
However, \cite{lsg94} found a number
of sdB stars in the super-metal-rich $(\rm [Fe/H] \sim 0.2;$ 
\cite{tbdh}) Galactic old open cluster NGC 6791. 
Very recently, Rich \etal\ (in preparation; see also Piotto \etal, these
proceedings)
found well-populated blue tails in the
metal-rich clusters NGC 6388 and NGC 6441. While the
process that gives rise to these hot stars is unknown (and may 
not be relevant to galaxies), taken with the field sdB/sdO objects,
these clusters provide an
``existence theorem'' for hot helium burning stars in metal-rich populations.

\subsection{The Nature of the UVX Phenomenon}

The most obvious originating stellar population for the UVX is
a small population of young, massive stars, and several studies based 
on this hypothesis appeared. However, later work with with the 
resolving power of HST (\cite{k93}) 
appears to rule out this hypothesis for M31. King \etal\
found an underlying smooth flux distribution increasing toward the nucleus 
with none of the discreteness found for OB associations, which in any
case had no concentration toward the centre. Also, some of the flux
($\sim 20\%$) emanated from resolved P-AGB stars.

Hills (1971) first suggested that post-asymptotic giant branch stars
might be the source of the upturn found in the bulge of M31
[$\xvv = 3.5$].  \cite{br90} and \cite{gr90} both showed that the maximum
possible UV flux given by the numbers of P-AGB stars expected from
standard stellar population models was insufficient to account for the
amplitude of the strongest UV upturns.
Ferguson \& Davidsen (1993)
contrasted Hopkins Ultraviolet Telescope (HUT) spectra of the
strongest UVX system, NGC1399 [\xvv\ = 2.05] and that of M31 and
concluded that (a) in the IUE
spectral range (\wmore{1250}) much of the flux emanated from sources with
characteristic temperature of $\sim 25,000 \kelvin$ and  (b) the spectral 
energy distributions
(SEDs) did not match at the short end of the spectrum (\wless{1200}),
with the flux from M31 being hotter.  This is consistent with the
notion that a larger proportion of the UV flux from M31 comes from the
P-AGB stars, which have a much higher time-averaged effective temperature. 
These conclusions are corroborated by the later studies of
\cite{bfd95} and \cite{bfdd}, who found similar characteristic
temperatures in six other elliptical/S0 galaxies.
Their spectral fits to the Astro-2 HUT data suggest,
 under a large range in assumptions about
the P-AGB stars in the galaxies, that at least some EHB stars are apparently
present in all systems. Thus the model that appears to be consistent
with the observations is of an old population where the UV flux
emanates partly from the P-AGB stars and partly from EHB stars.

\section{The Horizontal Branch Mass Dispersion}

The HB stars are the hottest potential source of UV radiation in
an old ($t \! \gta \! 2\, {\rm Gyr}$) stellar population
and have moderate intrinsic luminosities with appreciable stellar lifetimes.
However, it is well-known that modelling HB populations 
in Galactic globular clusters is subject to a major difficulty, 
{\it viz.} that of the mass dispersion. \cite{ir70} 
first realized that the observations implied a scatter in
HB properties, and \cite{rtr73} explicitly showed using
`synthetic' HB sequences that a dispersion in HB envelope
masses could reproduce the observations.
However this quantified rather than explained the dispersion in mass,
likely due to mass loss on the red-giant branch, by 
mechanisms still poorly understood to this day.
The hottest HB stars are produced by extreme mass loss 
(Ciardullo \& Demarque 1977; \cite{c89}; \cite{ct91}; DRO93).
Also, as clusters
age, the models imply that the RGB mass decreases so that
less mass needs to be stripped to produce hot HB stars in older
populations. Synthetic HB models thus show
a passive drift of the HB sequences blueward as a cluster ages, which 
mimics the effect of higher mass loss.

Iben \& Rood (1970) also noted the variations of HB stellar properties with
envelope mass, metallicity, helium abundance $Y$ (see also
\cite{sg76}), the CNO element fraction, as well as the core mass,
although the variations in the latter are constrained by energy arguments. 
As a result,  cluster observations  are subject
to many possible interpretations (aside from observational problems
such as determinations of reddening, calibration of blue stars, etc
etc).

In Galactic clusters it is possible to identify the resolved hot stars
as members of the HB.  In studying the stellar populations of galaxies,
however, we must start by asking a more basic  question, i.e. `which
stars are actually responsible for the UV radiation?' 
The following step is to explain the observed UVX
correlation in terms of the galaxy properties, but the interpretation
is still not straightforward (see \S 5).

In a detailed study of the possible explanations for the UVX, \cite{gr90}
suggested that an increase in the mass loss with metallicity
would produce blue HB stars. 
The assumptions involved both an increase in mass loss
with metallicity (almost universally acknowledged to happen, although
without quantitative theoretical backing), and an extrapolation of the
helium enhancement vs. metallicity relations derived at low metallicity
to the super-solar metallicity regime.
By using simple analytical fits to the
behavior of the HB stars and the HB morphology with increasing metallicity,
they showed how the UVX-metallicity correlation might be produced
naturally, by properties of stellar evolution rather than
of the galaxy environment. Environmental effects such as 
galaxy size, density or previous history are necessarily
more difficult to quantify, but may  be relevant to the problem:
we regard
these as  `stellar populations' inputs to the problem as distinct
from those of stellar evolution ({\sl cf.} \cite{d96}).

\section{Modelling of Ultraviolet Bright Old Populations}

The unresolved stellar populations of galaxies are studied using the 
methods of evolutionary population synthesis ({\it cf.} \cite{t80}; 
Renzini \& Buzzoni 1986; see also \cite{bc93}). 
Evolutionary population synthesis attempts to model the integrated
spectrum of a stellar population by summing the contributions at 
each individual wavelength arising from the population components
assumed to be present. We sketch here the derivation of
 values for UV/Optical colours such
as $\xvv.$

\subsection{Population Synthesis in a Nutshell}

Since the evolutionary connection between the early and late stages
of evolution is broken by the indeterminate nature of the RGB mass
loss process it is convenient to
break the problem into pre-He flash $\Lbando$ and post-He flash $\Lbandl$
stages.
We consider the integrated light
from each component with fixed composition ${\bf X} = (Y,Z_\alpha,Z_{\rm 
CNO},
Z_{\rm Fe-peak}$) etc.
For populations old enough to produce RGB sequences, the equation
for the luminosity in any given bandpass designated by $\lambda$
is (see DOR95 for more details)

\begin{eqnarray}
\protect{\Lband({\bf X})} & = &    \protect\Lbando + \protect\Lbandl 
\nonumber \\
   & = &  \int d\tau \int d\mass \lband(\mass;{\bf X})\Psi(\mass,{\bf 
X},\tau)d\mass  \nonumber \\
    & ~ & + \sbprate\Lvo\Delta V_0 
        \int_0^{M_{\rm RGB}-\mc} P(\mass;{\bf X})
        \eband(\mass;{\bf X})\, d\mass.
\label{popsyn}        
\end{eqnarray}

\noindent where

\begin{equation}
\eband(M,{\bf X}) = \int^{WD}_{ZAHB}{\lband(M,{\bf X})}dt.
\label{lifetime}
\end{equation}

The first term here is the contribution of the early (pre-RGB) evolutionary 
phases, and is shown here schematically as an integral over time and mass.
The time integral is over the interval where the star formation rate is 
significant, and the mass integral goes from the lower end of the main 
sequence to the RGB tip.
The luminosity \lband\ of a star at $\lambda$ is modelled using the relation

\begin{equation}
\lband(M,t,{\bf X}) = \bclam(g,\teff,{\bf X})\lbol(M,t,{\bf X})
\end{equation}

\noindent where $\bclam$ is the `bolometric correction,' i.e. the proportion
of the
total flux radiated in the band designated by $\lambda.$ This is quite
general: $\lambda$ can be a narrow pass (a few angstroms)
 for the construction of synthetic spectral energy distributions, or
 a broadband color such as a Johnson filter. The values of $\bclam$ are
 supplied using either an empirical flux library (more realistic, but
 never complete and hard to calibrate over long wavelength baselines), or
 synthetic stellar fluxes (e.g. \cite{k91}). The latter have the
 great advantage of completeness, but their accuracy particularly at
 short wavelengths is difficult to test at different metallicities.
 Their use also presupposes the accuracy of temperature/SED relations
 derived from the library of model stellar fluxes.

As for the late stages, the terms in the equation are the integrated
lifetime energy output from the ZAHB to the end of the white dwarf
stage. This term is simplified by the fact that the core mass
changes almost negligibly with cluster age (for $t \; \ga \; 5\, {\rm
Gyr}$),
so that the properties of the ZAHB are invariant with time.
The unknown mass distribution of the ZAHB stars due to variable mass
loss on the RGB is denoted by $P(m,t,{\bf X}).$
The lifetime energy flux [equation~(\ref{lifetime})] is multiplied by the `specific evolutionary flux,' denoted as
$\dot n_0(t,{\bf X}),$ which is the input rate {\it per unit magnitude
of pre-HB population} arriving on the HB. $\dot n_0$ was approximated by
Renzini \& Buzzoni (1986) and DOR95 as the number of stars leaving
the main sequence, but it is also possible to derive this number from
the RGB evolutionary tracks or empirically (\cite{rwo80}).
\Lvo\ is the $V$ band
luminosity from the pre-He flash stages, and $\Delta V_0$ is a constant, the
width of the $V$ filter $( = 872 {\rm \AA}).$

Potentially, both the IMF $\Psi$ and the HB mass
distribution function $P$ are affected by factors beyond stellar
evolution in single stars: this is the `stellar populations' component
to the problem. The effects of the initial conditions (such as rotation,
composition inhomogeneities, dynamics and environment) are not understood.
In the case of the IMF, we can show that for low mass stars, the
power law functions fit to cluster and field populations do not
much affect the output model spectra. The same is not true for the HB
mass distribution where small changes can produce large effects
in the ultraviolet output. It is thus somewhat misleading to
write $P(M)$ as an implicitly single-valued function of age
and composition, as would be correct if only stellar evolution
had an impact on the problem.
$P$ can however be constrained empirically and
the model tested for consistency against its other observational
predictions.

\begin{figure} 
\vspace{4.0in}
\includegraphics{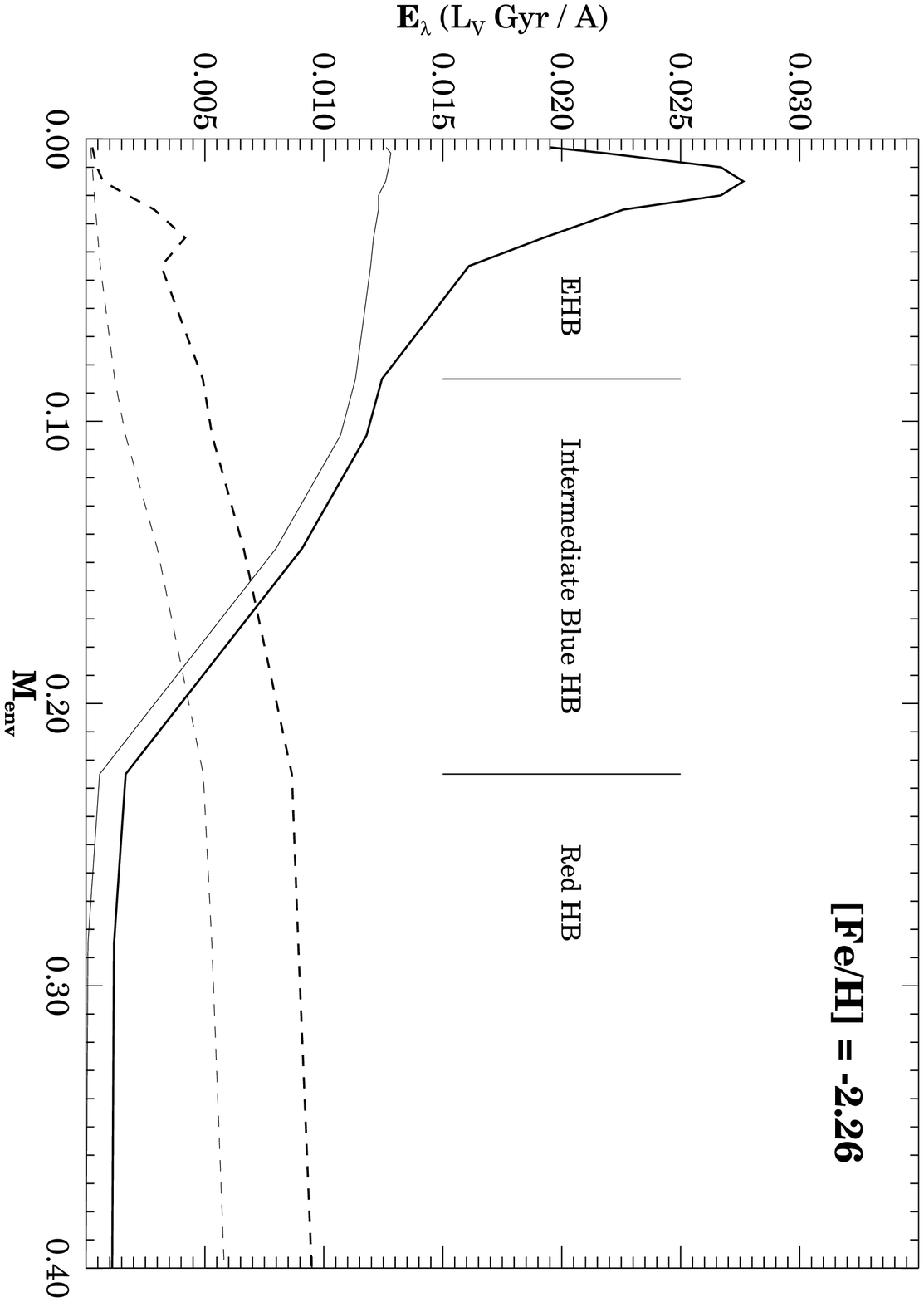}
\caption{The total energy radiated (\eband) by HB stars at $1500\; 
{\rm\AA},$
 and in the $V$ band as a function of envelope mass for 
 $\rm [Fe/H] = -2.26.$ The vertical axis is in
units of the solar $V$ band luminosity/Gyr/\AA. 
The data has been derived by integrating the
DRO93 HB evolutionary tracks and using the \protect\cite{k91} model
stellar fluxes. For the more massive stars, the $1500\; {\rm\AA}$ flux
is radiated entirely in the P-AGB phase of evolution, and the flux from
the lowest mass \protect\cite{sch83} evolutionary track has been
adopted (this gives an upper bound to the P-AGB flux; see DRO93).}
\label{eband}
\end{figure}

\subsection{The Energy Function \eband(M)}

Figure~\ref{eband} shows the lifetime energy radiated by HB stars as a
function of envelope mass at 1500 \AA\ and in the $V$ band.  Here we
have assumed that red HB stars evolve through the P-AGB channel that is
the most favourable (i.e., lowest mass) for UV production.  
The models used are from
DRO93 with the P-AGB flux from the \cite{sch83} $M_{bol} = -3.2$ $(M =
0.546\msun)$ model.
The thin lines show the energy radiated up to the core He
exhaustion point. The regimes of $\menv$ labelled show (a) red HB stars
(with far-UV radiation only in the P-AGB stage); (b) the IBHB stars,
 which reach
the AGB after core helium exhaustion: these are the stars that populate
the part of the HB usually thought of as the `blue HB' in globular
clusters such as M3; and (c)  the EHB stars whose total
post-HB energy radiated is similar to that emitted before core He
exhaustion.  The pre--core-He exhaustion energy radiated by the EHB
 stars is seen to be approximately constant with envelope mass and
 similar to the maximum post-HB output.
Models for other metallicities (DRO93, DOR95) show that  the
total far-UV radiation from all EHB stars from $\rm (Y,[Fe/H])
=(0.25,-2.26)$ to $(0.46, 0.58)$ varies by less than 50\%. Also, metal-rich
compositions produce the IBHB stars for only a small range in mass
(DRO93, \cite{ddro}), which is the primary cause for the `first
parameter' effect in globular clusters. This
also means that far-UV radiation from purely metal-rich populations
(above what can be produced by P-AGB stars) requires  mass loss in RGB
stars to be sufficiently vigorous to strip (almost) the entire
envelope.

The other factor that
influences the UV production from a stellar population is the rate of
evolution $\sbprate$ which is primarily a function of $Y,$ being
roughly double the rate at $Y_{\rm ZAMS} = 0.45$ as it is at $Y_{\rm ZAMS} =
0.27.$ This reflects the faster evolution of helium rich models.

\subsection{Simple Models of UV Populations}

Equation~\ref{popsyn} can be simplified to obtain some important bounds on
the UV-radiating population. Here, we focus on the number of hot 
stars that must be present for consistency with the observed far-UV 
radiation.
We assume a simple mass
distribution, initial mass function and metallicity distribution as follows.
The model consists of:

\begin{itemize}
\item an instantaneous starburst at a single metallicity $\Psi = \psi(M)$
\item a bimodal HB mass distribution with some fraction \fuv\ of the stars
becoming hot.
\end{itemize}

With the first assumption, our conclusions will of course not give precise
metallicity and age information but will instead give luminosity-weighted 
mean quantities. This  problem is more relevant to the study of
radiation in the mid-UV (Worthey, Dorman \& Jones 1996).

The HB mass distribution used is justified as follows: the red HB stars
of metal-rich populations  enter the integrated light mainly in the
$V$-band where their total energy output varies little with envelope
mass (this result can also be seen in the metal poor models of
Fig.~\ref{eband}, and explicitly in DOR95). The fraction  \fuv\
making up the EHB UV  emitters realistically is spread over a finite
mass distribution (i.e.  $\fuv = \int_{\rm \sst EHB} P(\mass) d\mass$).
This mass distribution cannot be easily
distentangled from integrated light, although \cite{bfdd}  find that a
small mass range of EHB stars produces the best models for the HUT
far-UV spectra. In order to avoid adding parameters to the problem that
cannot be constrained DOR95 used either single evolutionary sequences
and uniform distributions of EHB stars
 to model the hot contribution. The `single evolutionary sequence'
 models can provide an lower bound to the size of the
 hot population present, since a realistic distribution always produces a
 redder colour than the hottest stars present. Thus we may choose the
 peak of the  $\eband(M)$
 function to derive a bound. 
 
 Magnitudes and UV/Optical colours may now
 be derived from Equation~\ref{popsyn}. The result is

\begin{equation}
m_\lambda - V  = -2.5\log_{10}\Biggl(
{
{ (\Lband / \Lv)^E + \sbprate\Delta V_0 [ \fuv \ELhot + (1 - \fuv)\ELcool] }
 \over 
 { 1 + \sbprate\Delta V_0 [ \fuv \EVhot + (1 - \fuv) \EVcool] } }\Biggr).
\end{equation}

\noindent where

\begin{equation}
\ELhot = \int_{\rm EHB\; masses} \eband P(m^\prime) dm^\prime,
\end{equation}

\noindent and similarly \ELcool is the energy output integrated over
the red HB stars. Thus for any given observed  $\FUV = 10^{\sst [ -
0.4\xvv]},$
we obtain (since ${\cal L}_{\sst 1500}^{E} \approx 0)$

\begin{equation}
\fuv =  {{{\FUV(1 + \sbprate\Delta V_0 \EVcool)- \sbprate\Delta V_0\Exvcool}}
\over{\sbprate\Delta V_0\big[\Exvhot - \Exvcool+\FUV(\EVcool-\EVhot)\big]}}
\end{equation}

For the entire galaxy population, which is composite in abundance, we need
to use in this formula the average specific evolutionary flux 
$\overline\sbprate.$ A simple but  accurate approximation is

\begin{equation}
\fuv = { {\FUV}\over\overline\sbprate\Delta V_0\big[\Exvhot -
\Exvcool+\FUV(\EVcool-\EVhot)\big]}
\label{final}
\end{equation}

For the specific case where the mass function is taken to be the
Salpeter law, $\psi(M) = M^{-2.35}$, DOR95 have tabulated the
quantities  $(\Lband / \Lv)^E$ and $\sbprate,$ and also the values of
$E_{\rm 1500}$ and $E_V$ for tracks computed by DRO93. Apart from
the constant $\Delta V_0,$ the terms in the denominator of
equation~(\ref{final}) are the mean input rate to the HB and
the (mean) lifetime energy radiated by the hot stars.

\section{Discussion}

We have focussed here on deriving limits on the population size responsible for
the UV upturn. This is an important prediction of the theory, since is 
is the most easily tested for consistency against other observational
consequences. In particular, for the case of NGC 1399 [$\xvv = 2.05$] 
we obtain $\fuv \sim 0.16.$ This conclusion implies that the UVX sources
do not arise from a trace population more prevalent in
UVX-strong galaxies, and instead favours the notion that a
sizeable
minority of the dominant population is contributing to the UVX. 

The work of \cite{pl97} and \cite{bcf94} have assumed that a fraction of 
stars at the low and high end of the metallicity distribution contributes the
UV radiation from the galaxies. In the case of the metal-poor hypothesis, a 16\%
fraction of the population in stars with $\rm [Fe/H] < -1$ would give a large
enough contribution to the mid-ultraviolet radiation to produce a discrepancy
with observation (\cite{wdj96}; Bressan \etal\ 1994). Quantitative estimates employing
the 2500 \AA\ fluxes in DOR95 in equation~(\ref{final}) show that bimodal
metallicity distributions, in which the majority population
contributes relatively little in the mid-ultraviolet,
are necessary to reconcile that longer wavelength
flux with observations.

The hypothesis that the UVX stars are entirely metal-rich clearly does
not suffer from this problem.  However, the large fraction of stars
necessary to account for the observed flux seems to imply that a spread in
metallicity must give rise to the UVX rather than the most extreme
composition. This is contrary to the work of Bressan \etal\ who suggest
that the UVX arises from the effects of stellar evolution at large ages.
They invoke high metallicity, strongly helium enhanced models with RGB mass loss
 similar to what is inferred from the globular cluster system.

Assuming that it is a fraction of the
 dominant population that gives rise to the UVX accords  better with
the observational record we have for Galactic populations.
 We do not need to hypothesise that {\it all} RGB stars of a given 
metallicity have the same degree of mass loss, which we do not
observe in any other context (little information about mass distributions
can be deduced from red HB clumps).
Unfortunately one cannot test the initial metallicity of the
field sdB stars easily since their spectra are affected by diffusion
in the surface layers.
However, the question of how  the UVX arises remains open,  
at least for the time being (but see \cite{d97a} for empirical uses of 
the far-UV radiation).

 It should be stressed
 that both `metal-poor' and `metal-rich' models
 as presented by these authors rely on the hypothesis
 that age is established as a `global' second parameter
 for HB morphology in old stellar systems.
 They use the results of simple, unimodal synthetic HB models to
 derive ages of galaxies. We argue that the Galactic record in
 resolved populations does not support the use of these models
 and cannot therefore constrain unresolved stellar populations.

 We close with a caveat that applies to the interpretation
 of the UVX as a metallicity-driven phenomenon, on which
 the `metal-rich' hypothesis relies heavily.
 We caution that this assumption should not be regarded as
 incontrovertible. The argument for a metallicity driven mechanism
 is weakened by the discovery that the \mgii\ indicator
 does not trace the heavy-element abundance in galaxies 
 and by the observation that \mgii\ line strengths are correlated with 
 $\sigma_0$ but not with iron indicators (\cite{wfg92}).
 The UVX may thus be uncorrelated with iron abundance
 (\cite{d97a} and references therein).
 Another definite prediction of the metallicity
 relation is that the UVX gradient with galactocentric radius should
 be correlated with optical line-index gradients. This question
 is currently under study using Ultraviolet Imaging Telescope data
 (Ohl \etal, in preparation).
 Thus, the Ultraviolet Upturn Phenomenon
 remains, like many others in astronomy, an unsolved problem
 which requires spatially resolved spaceborne observations for
 further study.

Acknowledgements: B.D. acknowledges support from NASA grants NAG5-700
and NAGW-4106, and healthy discussions with Robert T. Rood and Robert
W. O'Connell. He is also glad to report that Icko Iben Jr., who we honour
with this volume, did not find anything wrong with his presentation.

{} 

\end{document}